# Spiral Structure In the Circumnuclear Disk At The Center Of NGC 4258

Eyal Maoz

Astronomy Dept., University of California at Berkeley, CA 94720*

E-mail: maoz@astro.berkeley.edu

* Miller Fellow.

## ABSTRACT

Observations of line emission from water masers near the center of the galaxy NGC 4258 have recently provided compelling evidence for rotating disk of gas, viewed nearly edge-on, surrounding a massive black hole. We show that the disk is very likely to be only marginally stable to radial perturbations - a stability regime where weak non-axisymmetric disturbances grow via the "swing amplification" effect, leading to the formation of a ragged, multi-armed spiral pattern similar to that observed in Sc galaxies. This suggests a natural explanation for the apparent clustering of the high-velocity emission sources into several distinct clumps, and for the observed regularity in the distance intervals between them. The clumps of maser sources appear at the intersections of the spiral arms and the radial line of longest coherent gain path (the diameter through the disk perpendicular to the line-of-sight), and are thus spaced apart at the characteristic crest-to-crest radial distance between the arms.

This interpretation implies a disk thickness of $\simeq 0.003\,(\bar{Q}/1.6)\,\mathrm{pc}$ at a radius of $\simeq 0.2\,\mathrm{pc}$, where the local value of the stability parameter is $1.2 \lesssim \bar{Q} \lesssim 2$. The $H_2$ density is $\simeq 1.8 \times 10^{10}\,(\bar{Q}/1.6)^{-1}\,\mathrm{cm}^{-3}$, assuming that $H_2$ dominates the mass density in that region. The disk mass is $\lesssim 10^6 M_\odot$, which is consistent with the accuracy of the Keplerian fit to the rotation curve of the maser emission sources. These results closely concur with previous estimates based on independent considerations.

## 1. INTRODUCTION

The maser line emission from the circumnuclear disk in NGC 4258 consists of two groups of sources (Miyoshi *et al.* 1995). The systemic features appear along the line-of-sight to the center of the disk and have velocities comparable to the systemic velocity of the galaxy. The high-velocity features are distributed in a nearly planar structure at distances between 0.16 pc (hereafter $r_{in}$) and 0.25 pc ($r_{out}$) from the center of rotation, and their line-of-sight velocities are described by Keplerian motion to high precision. This remarkable finding confirmed earlier suggestions (Watson & Wallin 1994) that the high-velocity sources lie along the diameter through the disk (mid-line), perpendicular to the line-of-sight. The restricted spatial distribution of masers along the line-of-sight to the nucleus and along the mid-line of the disk are undoubtedly related to the fact that in these regions the gradient of the line-of-sight velocity is zero and hence, these regions support long path-lenghts for coherent amplification in *our direction* (Miyoshi *et al.* 1995). Obviously, all observers in the universe who view the disk nearly edge-on would see a qualitatively similar picture.

The high-velocity maser sources are not distributed uniformly nor randomly along the mid-line. They are clustered into seven distinct "clumps," five on one side of the disk, and two on the other (figure 2a in Miyoshi *et al.* [1995]). Moreover, there seem to be some regularity in the distance intervals between the clumps; they are almost equally spaced apart from each other, with mean inter-clump separation of $\simeq 0.023$ pc (0.75 mas). It is plausible that the intriguing clustering of sources is due to small gas clumps in the disk inside which the conditions for amplification of maser emission are favorable for as yet unknown reason, but this simple interpretation is improbable for the following reason: if noticeable maser emission had been produced due to amplification over the column density within a gas clump, the chances of detecting that emission would have been independent of the location of the gas clump in the disk. In such case, we should have detected emission from gas clumps also at locations other than along the mid-line - a situation which is firmly ruled out by the observed nearly perfect Keplerian rotation curve.

We shall now argue that the observed clustering of the maser sources, and the indicated regularity in the distance intervals between them can be naturally explained by simple disk stability considerations (§2). In §3 we conclude and discuss caveats.

## 2. DISK STABILITY AND SPIRAL STRUCTURE

### 2.1. The Condition For Spiral Structure

A rotating, self-gravitating cold disk is locally stable against axisymmetric disturbances of wavelength $\lambda > \lambda_{crit}$, where

$$\lambda_{crit} \equiv \frac{4\pi^2 G\Sigma}{\kappa^2} \quad , \tag{1}$$

$\Sigma$ is the surface density of mass per unit area, $\kappa(r) \equiv \Omega\,[4 + d\log(\Omega^2)/d\log r]^{1/2}$ is the local epicycle frequency, and $\Omega(r)$ is the angular velocity. A non-rotating sheet with non-zero sound speed, $v_s$, is stable against perturbations with $\lambda < v_s^2/G\Sigma$. Thus, under the joint influence of rotation and moderate superposed sound speed, the unstable axisymmetric disturbances are confined to a certain intermediate range of scales - a range that shrinks as the sound speed is increased. A thin gaseous disk is locally stable to all axisymmetric perturbations when $Q > 1$, where (Toomre 1964; Julian & Toomre 1966)

$$Q \equiv \frac{v_s \kappa}{\pi G \Sigma} \quad . \tag{2}$$

Although $Q > 1$ curbs the axisymmetric Jeans instabilities, a considerable growth of non-axisymmetric disturbances occurs when $1 < Q \lesssim 2$, leading to the formation of a multi-armed spiral density pattern (Goldreich & Lynden-Bell 1964; Julian & Toomre 1966; Toomre 1981, 1990). This phenomenon was discovered in the context of spiral galaxies (Goldreich & Lynden-Bell; Julian & Toomre 1966), and is believed to be the basic engine behind the structure of Sc galaxies. In contrast to "grand design" spirals like M51 or M81, Sc galaxies exhibit a ragged pattern of many spiral arms. The underlying physical mechanism which is responsible for this structure is the "swing amplification." It results from a three-fold conspiracy between the shear flow, epicyclic shaking and self-gravity (Goldreich & Lynden-Bell 1964; Julian & Toomre 1966). The disk particles travel almost freely through the density pattern, so any individual spiral arm does not remain composed always of the same particles. The spiral pattern is constantly drifting and shearing with the differential rotation, and spiral arms are constantly forming and dying with typical life-time of order half the revolution period. As observed in simulations (Miller, Prendergast & Quirk 1970; Hohl 1970; Hockney & Brownrigg 1974; Toomre 1981; Sellwood & Carlberg 1984; Carlberg & Freedman 1985), even weak transient perturbations with relative force of a fraction of a percent can yield an evolving spiral pattern of impressive severity. The maximum growth occurs for waves of *statistically* well-defined wavelength (Goldreich & Lynden-Bell 1964; Julian & Toomre 1966), leading to a spiral structure which is most pronounced at radial crest-to-crest separation of $\simeq \lambda_{crit}$, as defined in equation (1). The less self-gravitating the disk, the smaller is the radial distance between the arms, and the larger is the number of arms (*e.g.*, Carlberg & Freedman 1985).

### 2.2. The Circumnuclear Disk

We suggest that the stability state of the circumnuclear disk in NGC 4258 is similar to that of Sc galaxies, implying that it too may well have a ragged, multi-armed spiral structure within some range of radii. This could explain the apparent clustering of maser emission sources as appearing where spiral arms intersect the mid-line (Figure 1). Since these intersections are spaced $\simeq \lambda_{crit}$ apart, it would produce some regularity in the distance intervals between the clumps, as observed. We shall now derive the disk characteristics required for the existence of a spiral structure at the region of the high-velocity sources.

The mean distance of the high-velocity sources from the center of rotation is $\bar{r}\simeq 0.20\,\mathrm{pc}$. The thickness of the disk, $\bar{h}$, at radius $\bar{r}$, and the local circular velocity, $\bar{v}$, are related by the condition of an hydrostatic equilibrium which reads $\bar{h}/\bar{r}\simeq\bar{v}_s/\bar{v}$, where $\bar{v}_s$ is the effective sound speed in the disk at radius $\bar{r}$ (notice that $\bar{v}_s$ need not necessarily be directly related to the temperature of the masering gas, *e.g.*, in the case of the interstellar medium in galactic disks the internal temperature of diffuse $H_I$ clouds is significantly lower than the root mean-square random cloud velocity [Spitzer 1978]). Since the force field is Keplerian to high precision (Miyoshi *et al.* 1995) we have $\kappa(\bar{r}) = \Omega(\bar{r}) = \bar{v}/\bar{r} = 1.42\times 10^{-10}\,\mathrm{s}^{-1}$. Combining the above equalities with equations (1) and (2) we obtain

$$\bar{h} \;=\; \frac{\bar{Q}\lambda_{crit}}{4\pi} \;\simeq\; 0.1\left(\bar{Q}/1.6\right)\;\;\mathrm{mas}\quad , \tag{3}$$

where $1.2 \lesssim \bar{Q} \lesssim 2$ is required for the existence of a spiral structure, and $\lambda_{crit}\simeq 0.75\,\mathrm{mas}$ is the mean radial distance between the clumps. Assuming that $H_2$ makes the dominant contribution to the mass density of the disk at radius $\bar{r}$, and that the relative fraction of water vapor is negligible (*e.g.*, Anderson & Watson 1993), the density of $H_2$ molecules at $\bar{r}$, averaged over the thickness of the disk, is

$$\bar{n}_{\mathrm{H}_2} \;=\; \frac{\Omega^2(\bar{r})}{\pi G m_{\mathrm{H}_2}\bar{Q}} \;\simeq\; 1.8\times 10^{10}\left(\bar{Q}/1.6\right)^{-1}\;\;\mathrm{cm}^{-3}\quad . \tag{4}$$

For comparison, Miyoshi *et al.* (1995) estimated a disk thickness of $<0.1\,\mathrm{mas}$ and $H_2$ density of $\sim 10^{10}\,\mathrm{cm}^{-3}$ based on independent considerations.

Since the disk is not massless it must introduce some systematic deviation from a perfect Keplerian rotation curve. Although the surface mass density profile of the disk is unknown, the above derivation fixed its value at radius $\simeq\bar{r}$ ($\bar{\Sigma}=\bar{n}\bar{h}m_{\mathrm{H}_2}$). Let us assume, for example, that $\Sigma(r)=\bar{\Sigma}\bar{r}/r$. The rotation profile in the absence of the black hole would have been flat, with $v_c^2=2\pi G\bar{\Sigma}\bar{r}$. Thus, the rotation curve in the combined force field is $v^2(r) = (\alpha/r)+v_c^2$, rather than $v'^2(r)\equiv\alpha'/r$. The systematic deviation of $v(r)$ from $v'(r)$ over the entire range of radii $[r_{in}, r_{out}]$ is given by $\Delta v \simeq v(r_{in}) - v'(r_{in})$ under the constraint that $v(r_{out})=v'(r_{out})$. We obtain

$$\Delta v \;\simeq\; -4.2\left(\frac{\bar{n}_{\mathrm{H}_2}}{1.8\times 10^{10}\,\mathrm{cm}^{-3}}\right)\left(\frac{\bar{h}}{0.1\,\mathrm{mas}}\right)\,\mathrm{km\,s}^{-1}\quad , \tag{5}$$

which is independent of $Q$. Such deviation is consistent with the accuracy of the Keplerian fit to the rotation curve (Jim Herrnstein, private communication). The minus sign implies that the rotation curve should be slightly flatter than an $r^{-1/2}$ power law. If $\Sigma(r)\propto r^{-1}$ all the way to the center, the total mass of the disk within $r_{out}$ is $\simeq 9\times 10^5\;(\bar{n}/1.8\times 10^{10}\,\mathrm{cm}^{-3})\left(\bar{h}/0.1\,\mathrm{mas}\right)M_\odot$. Assuming a steeper surface density profile at $\bar{r}$, in which case the associated rotation curve falls with distance, would result in even smaller deviation from a Keplerian relation.

## 3. DISCUSSION

We have shown that the puzzling distribution of high-velocity maser emission sources can be attributed to a spiral structure in the circumnuclear disk. This interpretation implies a mean disk thickness (Eq. [3]) and density (Eq. [4]) which closely concur with previous estimates that are based on independent considerations. While spiral structure on scales of $\sim 10^4$ parsec in Sc galaxies is outlined by tracers such as $H_{II}$-regions, OB-star associations, 21-cm line emission, and CO lines, a similar spiral pattern on a sub-parsec scale appears to be outlined by maser line emission, thus providing unprecedented insight into the structure of circumnuclear disks.

The spiral structure interpretation implies that the conditions for amplification of maser emission are more favorable in the wave crests than in the troughs. This raises an interesting question: how large should the density contrast in the spiral pattern be, relative to the mean density of the disk, in order for the maser emission to be produced preferentially in the spirals arms?

Answering this question is not straightforward since the condition for the onset of masering, as well as the maser luminosity, depend not only on the density but also on the temperature, the $H_2O$ /$H_2$ ratio, the amount of dust grains (*e.g.*, Collison & Watson 1995), the nature of the pumping mechanism, and on whether the emission is saturated or not. There is currently no theory which gives the maser luminosity as a function of all these factors (M. Elitzur, C. McKee & D. Hollenbach, private communication), except in a few specific cases. For example, assuming that the pumping mechanism is collisional excitation, which is likely to be the case for the high-velocity masers, the brightness temperature of the maser emission has been shown to scale as $\rho_{gas}^2$, where $\rho_{gas}$ is the gas density (Elitzur, Hollenbach & McKee 1989, 1995 in preparation). Given this non-linear dependence on density, and the fact that the S/N ratio for the intensity of the high-velocity features is $\lesssim 10$ (Figure 1 of Miyoshi *et al.* 1995), it is clear that a density ratio of $\rho_{crest}/\rho_{trough} \lesssim 3$ would be enough to reproduce the observations (notice that $\rho_{crest}/\bar{\rho}_{disk} < \rho_{crest}/\rho_{trough}$). A density contrast of order unity can be easily attained by a spiral pattern. It has been shown (Toomre 1981, and references therein) that the density contrast of a spiral arm depends on the local value of the Q parameter, and it changes in time as the arms shear, wind up, and disperse. The *maximum* amplification factor varies between $\sim 3$ when $Q \simeq 2$, to $\sim 100$ when $Q \simeq 1.2$.

In fact, the observed distribution of maser sources can be reproduced even by a lower density contrast than that suggested in the above discussion. The gas temperature in the spiral arms is likely to be higher than that between the arms, and it is known that the brightness temperature increases (in a non-trivial way) with temperature (D. Hollenbach, private communication). Studying the dependence of the maser luminosity on all the potentially important factors is far beyond the scope of this *Letter* and is left for a future investigation. Again, we emphasis that the condition for spiral activity and the radial distance between the arms are independent of the density contrast.

The existence of a spiral structure in the disk implies that the velocity field must deviate to some degree from a smooth Keplerian rotation curve. The amplitude of the deviations from circular velocities (hereafter, the peculiar velocity) is comparable to the amplitude of the epicyclic velocity around the guiding center, and it can be estimated in the following way: the orbital radius of the epicyclic motion is $\simeq \lambda_{crit}/2$, where $\lambda_{crit}$ is the mean distance between the arms. In the force field of a point mass the epicyclic frequency is equal to the orbital frequency of the guiding center and we have $v_{epicyclic} \simeq (\bar{v}\lambda_{crit}/2\bar{r}) \simeq 45\mathrm{km\,s^{-1}}$, where $\bar{v}$ is the orbital velocity at radius $\bar{r}$. Although such velocities are larger than the measured accuracy of the Keplerian rotation curve ($\sim 4\,\mathrm{km\,s^{-1}}$), we should not be able to detect these motions for a simple reason: at any point in the disk the local peculiar acceleration due to the spiral pattern is always orthogonal to the wave troughs or crests (Toomre 1981), and the peculiar velocities are predominantly along the same direction. Since the high-velocity maser emission is observed only along diameters through the disk which do not deviate from the mid-line by more than $6^\circ \simeq 0.1\,\mathrm{rad}$ (L. Greenhill, private communication), the peculiar velocities are almost perpendicular to the line-of-sight and are thus almost undetectable (the component of the peculiar velocity along the line-of-sight would not exceed $\simeq 0.1{\times}45 = 4.5\mathrm{km\,s^{-1}}$). Since the detected deviations from a Keplerian rotation curve are not much larger than the above value, we can also conclude that the pitch angle of the spiral arms, within the observed range of radii, does not exceed $\simeq 6^\circ$.

Explaining the difference in the distribution and intensity of red-shifted versus blue-shifted sources remains a serious challenge for theory. It is currently hard to understand this asymmetry, regardless of whether there is a spiral structure or not. It might be related to the fact that the disk is slightly warped (see Neufeld & Meloney 1995; Miyoshi *et al.* 1995).

We should emphasize that finding the disk near the border of stability ($1 < Q \lesssim 2$) does not necessarily requires fine tuning (Goldreich & Lynden-Bell 1964; Julian & Toomre 1966). Although the individual disk particles travel almost freely through the wave pattern, the constantly-evolving spirals tend to slowly increase the gas temperature, thus increasing the Q-parameter. In order to maintain a spiral activity the disk must be cooled. In Sc galaxies, energy dissipation in small regions in the gas leads to the formation of condensations such as giant molecular clouds, a process which effectively cools down the disk. The non-axisymmetric disturbances associated with the new condensations trigger new modes, and generation after generation of spiral arms form, wind up, and disperse. This has been suggested to be the self-regulating, regenerative mechanism of spiral structure in Sc's, and it may as well apply to the circumnuclear disk in NGC 4258. If the temperature of the disk is indeed maintained by this process, the implied slow depletion of gas should be balanced by continuous gas accretion.

Finally, we point out an alternative conceivable explanation for the clustering of the high-velocity sources. It has been shown (Syer, Clarke & Rees 1991, and references therein) that the cumulative effect of many passages of a star through the disk would bring it into a circular orbit corotating with the disk. If the disk is thinner than the star's Roche radius, the star may open a gap in the disk. The clustering of the maser sources could then be a manifestation of the ring-like structure of the disk due to the grooves produced by many such stars. However, we find this interpretation improbable mainly because the distance intervals between the clumps are substantially larger than the gravitational "sphere of influence" of a star in that disk.

I thank Alar Toomre, George Field, Jim Moran, Lincoln Greenhill, Jim Herrnstein, Charles Gammie, Moshe Elitzur, and David Hollenbach for discussions. This work was curried out at the Harvard-Smithsonian Center for Astrophysics, and was supported by the U.S. National Science Foundation, grant PHY-91-06678.

---

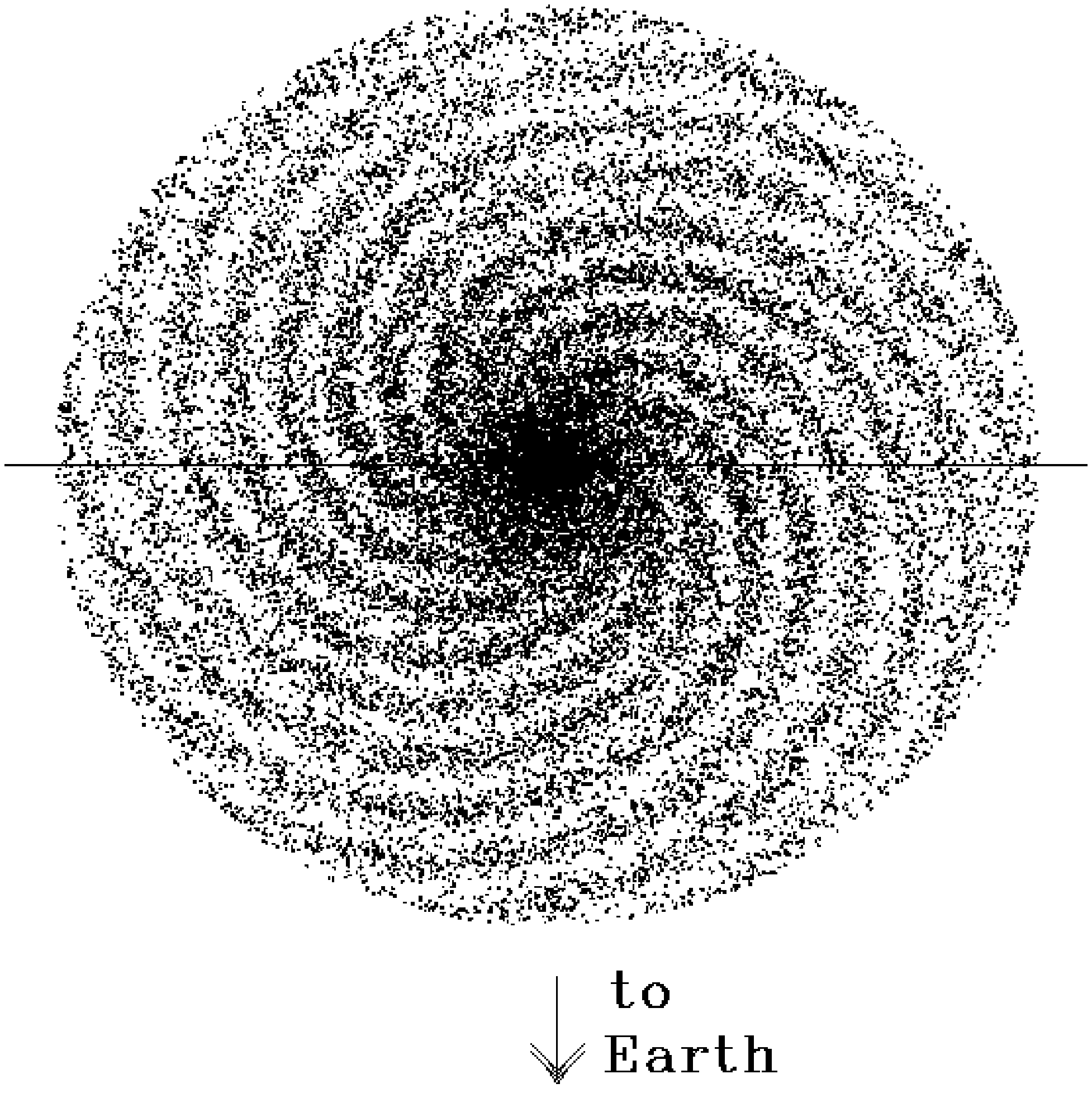

Fig. 1.— An illustration of a multi-armed spiral structure in the circumnuclear disk. The solid line represents the "mid-line" along which the coherent gain path for amplification of maser emission in our direction is long. It is argued that the clumps of high-velocity sources appear at the intersections of the mid-line and the spiral arms (see text). It should be emphasized that the real spiral pattern is unlikely to be as regular as in the above illustration. It probably resembles the structure of Sc galaxies which can be described as a swirling hotch-potch of pieces of spiral arms. The apparent deficiency in the number of blue shifted clumps may be attributed to the indicated large scale warp in the disk.